\documentclass[sigconf]{acmart}

\setcopyright{acmlicensed}
\copyrightyear{2018}
\acmYear{2026}
\acmDOI{XXXXXXX.XXXXXXX}
\acmConference[womENcourage'26]{ACM womENcourage}{September 30--02 October,
  2026}{Nice, France}
\acmISBN{978-1-4503-XXXX-X/2018/06}

\begin{document}

\title{Breaking In and Reaching Out: Networking for Women in Computer Science}

\author{Shalini Chakraborty}
\email{shalini.chakraborty@uni-bayreuth.de}
\orcid{0000-0002-9466-3766}
\affiliation{%
  \institution{University of Bayreuth}
  \city{Bayreuth}
  \country{Germany}
}

\renewcommand{\shortauthors}{Chakraborty}

\begin{abstract}
Networking is central to careers in computer science, where a globally distributed and diverse community increasingly collaborates across institutional and geographic boundaries, often in hybrid and remote settings. However, access to effective networking is shaped by structural and personal factors, including geography, funding, language, identity, personality, and caregiving responsibilities. Building on prior work, this workshop focuses on women in computing to examine lived experiences of networking and the barriers they encounter. Through a community-driven discussion grounded in a factor-based framework, the workshop aims to surface overlooked challenges and foster shared understanding. Ultimately, it seeks to inform more inclusive, equitable, and accessible networking practices within the computer science community.

\end{abstract}



\maketitle

\section{Introduction}
Networking plays a critical role in shaping research careers in computer science, spanning both academia and industry. It influences collaboration opportunities, visibility, mentorship, and long-term professional impact. However, as highlighted in \cite{chakraborty2026struggling}, networking is not merely an individual skill but a process deeply shaped by structural, social, and personal factors. These include geography, access to funding, immigration and regulatory constraints, language barriers, gender and social identity, personality traits, and personal responsibilities.

The role of professional networks in shaping academic careers has been well-documented across disciplines. Researchers with broader networks tend to collaborate more, publish more frequently, and receive greater citation impact \cite{barabasi2002new, crane1973invisible, filipovic2023social}. In SE, collaborations often emerge through conferences, workshops, and informal professional circles \cite{storey2010impact}. These venues function not only as spaces for sharing research but also as critical social infrastructures that define who participates in the field’s intellectual discourse.

Despite this importance, networking opportunities are unequally distributed. Prior work has highlighted the structural barriers faced by early-career researchers, women, and scholars from the Global South in accessing collaborative spaces and funding networks \cite{bilecen2017introduction, mokhachane2024voices, curry2017global}. Socioeconomic and geographic disparities affect who can afford to attend international conferences, while visa restrictions and institutional funding constraints disproportionately limit researchers from low- and middle-income countries \cite{gulel2025navigating}. 
Language and cultural differences further influence participation. English remains the dominant language of publication and communication, often placing additional burdens on non-native speakers \cite{lillis2011academic}. These linguistic hierarchies shape not only who can publish but also who feels comfortable engaging in informal networking settings such as conference receptions or online discussions.
In \cite{chakraborty2026struggling}, we argue that opportunities to build professional networks are unevenly distributed, often privileging individuals with greater access to resources, mobility, and established academic environments, while marginalizing others. These disparities are particularly evident in conference settings, where informal interactions—such as hallway conversations, social events, and community gatherings—play a central role in shaping collaborations and visibility.

Building on this work, the proposed workshop extends the discussion into an interactive, community-driven space with a specific focus on women in computing and related fields. While prior research has identified broad structural barriers, there remains a need to better understand how these challenges are experienced, negotiated, and perceived by women across diverse backgrounds and contexts.

This workshop is inspired by ongoing research on networking experiences in software engineering and computer science \cite{chakraborty2026struggling}. It builds on preliminary empirical evidence gathered through a survey conducted at ICSE'26, as well as follow-up qualitative and quantitative data collected within the Brazilian software engineering community via online social media platforms. \textbf{Please note that}, Rather than extending this study, the workshop leverages its insights as a motivation to bring together a diverse group of women in computing for an open and timely discussion on networking experiences. 

In particular, the opportunity to gather many remarkable women across computer science disciplines provides a valuable space to reflect on shared challenges and diverse perspectives. This is especially important for early-career researchers, for whom networking plays a decisive role in shaping academic pathways, collaborations, and visibility. Broadening the discussion beyond software engineering enables richer exchanges and may surface additional factors and nuanced experiences that are often overlooked in existing studies.

To guide the discussion, the workshop adopts and extends the factor-based framework introduced in the paper, focusing on the following dimensions:

\begin{itemize}
    \item Geography and Location
    \item Funding and Other Resources
    \item Nationality and Regulation
    \item Timing and Duration of Conferences
    \item Language and Communication
    \item Gender and Social Identity
    \item Personality and Neurodiversity
    \item Caregiving Status and Personal Responsibilities
\end{itemize}

Through these themes, the workshop aims to (i) understand women’s lived experiences and challenges in networking, (ii) identify overlooked or emerging factors, and (iii) contribute to a broader, community-driven effort to make networking in SE more inclusive, equitable, and accessible.

\section{Organizer and Presenter}

The workshop is organized and presented by Shalini Chakraborty. 

\textbf{Shalini Chakraborty} is a postdoctoral research associate and chair of the Software Engineering Group at University of Bayreuth. Her work lies at the intersection of technology, human factors and the evolving socio-technical practice of software engineering. With expertise in software engineering, model-based engineering, empirical research and human aspects, her research explores how software engineering practices are evolving beyond purely functional competencies toward more trustworthy, diverse, inclusive and ethical dimensions.

While the organizer will facilitate the session and introduce the research context and framework, the workshop is intentionally designed as a community-centered space.
Participants are encouraged to actively contribute, share experiences, and reflect on their own networking journeys. The goal is to move beyond a traditional top-down format and instead create a collaborative environment where every participant has the opportunity to voice their perspectives, challenges, and insights.

\section{Target Audience}

The target audience for this workshop is women participants of WoMenEncourage who are engaged in computer science, software engineering, and related fields, including academia, industry, and research.

The workshop is particularly relevant for:
\begin{itemize}
    \item Early-career researchers and students navigating networking spaces.
    \item Researchers and practitioners from diverse geographic and cultural backgrounds.
    \item Individuals interested in equity, diversity, and inclusion in computing.
\end{itemize}

While the primary focus is on women’s experiences, the workshop welcomes participants with diverse perspectives and encourages intersectional discussions across identities and contexts.

\section{Workshop Activities}

The workshop is designed as a 1.5-hour interactive session structured as follows:

\begin{itemize}
    \item \textbf{Introduction (15 minutes):} The organizer will present the motivation, summarize key insights from the ICSE-CHASE paper, and introduce the factor-based framework and ongoing survey.

    \item \textbf{Guided Discussion (60 minutes):} The session will be organized around the identified factors (e.g., geography, funding, language, gender, caregiving). For each category, participants will be invited to:
    \begin{itemize}
        \item Share personal experiences and challenges
        \item Reflect on how these factors influence their ability to network
        \item Identify gaps or additional factors not yet considered
    \end{itemize}

    The discussion will emphasize inclusivity, respect, and openness. No recording will take place to ensure a safe and supportive environment.

    \item \textbf{Reflection and Contribution (15 minutes):} Participants will be invited to optionally reflect on the discussion and share their thoughts if they wish. This activity is intended purely as a means of personal reflection and collective exchange, allowing participants to express their perspectives in a way that feels comfortable to them, without any expectation or obligation to contribute to research outcomes.

\end{itemize}

The workshop will prioritize creating a safe space for open dialogue. Participants may use laptops if they wish, but no technical setup is required. The focus is on conversation, reflection, and collective understanding rather than formal outputs.

Overall, the workshop aims to foster a sense of community, validate diverse experiences, and contribute to a broader effort toward more inclusive networking practices in software engineering.

\bibliographystyle{acm}
\bibliography{refs}

@article{chakraborty2026struggling,
  title={Struggling to Connect: A Researchers' Reflection on Networking in Software Engineering},
  author={Chakraborty, Shalini},
  journal={arXiv preprint arXiv:2601.10907},
  year={2026}
}

@article{barabasi2002new,
  title={The new science of networks},
  author={Barab{\'a}si, Albert-L{\'a}szl{\'o}},
  journal={Cambridge MA. Perseus},
  year={2002}
}

@misc{crane1973invisible,
  title={Invisible colleges: Diffusion of knowledge in scientific communities},
  author={Crane, Diana and Kaplan, Norman},
  year={1973},
  publisher={American Institute of Physics}
}

@article{filipovic2023social,
  title={Social capital theory perspective on the role of academic social networking sites},
  author={Filipovic, Jelena and Arslanagic-Kalajdzic, Maja},
  journal={Journal of business research},
  volume={166},
  pages={114119},
  year={2023},
  publisher={Elsevier}
}

@inproceedings{storey2010impact,
  title={The impact of social media on software engineering practices and tools},
  author={Storey, Margaret-Anne and Treude, Christoph and Van Deursen, Arie and Cheng, Li-Te},
  booktitle={Proceedings of the FSE/SDP workshop on Future of software engineering research},
  pages={359--364},
  year={2010}
}

@misc{bilecen2017introduction,
  title={Introduction: international academic mobility and inequalities},
  author={Bilecen, Ba{\c{s}}ak and Van Mol, Christof},
  journal={Journal of ethnic and migration studies},
  volume={43},
  number={8},
  pages={1241--1255},
  year={2017},
  publisher={Taylor \& Francis}
}

@article{mokhachane2024voices,
  title={Voices of silence: experiences in disseminating scholarship as a global south researcher},
  author={Mokhachane, Mantoa and Green-Thompson, Lionel and Wyatt, Tasha R},
  journal={Teaching and Learning in Medicine},
  volume={36},
  number={2},
  pages={235--243},
  year={2024},
  publisher={Taylor \& Francis}
}

@book{curry2017global,
  title={Global academic publishing: Policies, perspectives and pedagogies},
  author={Curry, Mary Jane and Lillis, Theresa},
  volume={1},
  year={2017},
  publisher={Multilingual Matters}
}

@article{gulel2025navigating,
  title={Navigating visa inequities: mobility as privilege in academia--‘You are not supposed to be here’},
  author={G{\"u}lel, Devran},
  journal={Global Social Challenges Journal},
  volume={1},
  number={aop},
  pages={1--10},
  year={2025},
  publisher={Bristol University Press}
}

@article{lillis2011academic,
  title={Academic writing in a global context: The politics and practices of publishing in English},
  author={Lillis, Theresa and Curry, Mary Jane},
  journal={J Bus Tech Commun},
  volume={22},
  pages={179--198},
  year={2011}
}
\end{document}